# High-pressure synthesis and superconductivity of the novel Laves phase BaIr$_2$

Terunari Koshinuma[1,2,*], Hiroki Ninomiya[2], Izumi Hase[2], Hiroshi Fujihisa[2], Yoshito Gotoh[2], Kenji Kawashima[2,3], Shigeyuki Ishida[2], Yoshiyuki Yoshida[2], Hiroshi Eisaki[2], Taichiro Nishio[1,2], and Akira Iyo[2,*]

[1]Department of Physics, Graduate School of Science, Tokyo University of Science, Shinjuku, Tokyo 162-8601, Japan

[2]National Institute of Advanced Industrial Science and Technology (AIST), Tsukuba, Ibaraki 305-8568, Japan

[3]IMRA JAPAN Co., Ltd., Kariya, Aichi 448-8650, Japan

**ABSTRACT**

Superconductors comprising 5$d$ transition metals of Ir and Pt have been widely explored because they have the potential of unique superconductivity caused by the strong spin-orbit interaction (SOI). We successfully synthesized BaIr$_2$, the last Laves phase remaining unsynthesized in the MgCu$_2$-type $AM_2$ ($A$ = Ca, Sr, Ba; $M$ = Rh, Pd, Ir, Pt). BaIr$_2$ was crystallized at 925 °C under a pressure of 3.3 GPa via a solid-state reaction between Ba and Ir powders; it was found to have the longest $a$-lattice constant of 8.038(1) Å among $AM_2$. BaIr$_2$ exhibited bulk superconductivity at a transition temperature ($T_c$) of ~2.7 K. BaIr$_2$ was found to have a type-II superconductor with an upper critical field of 67.7 kOe, which was above the Pauli paramagnetic limit (~50 kOe). The electron–phonon coupling constant and normalized specific heat jump were measured to be 0.63 and 1.2, respectively, indicating that BaIr$_2$ is a weak-coupling superconductor. The electronic-structure calculations for BaIr$_2$ revealed that the Ir-5$d$ states are dominant at the Fermi energy ($E_F$) and the density of states at the $E_F$ is strongly affected by SOI as in the case of CaIr$_2$ and SrIr$_2$.

∗ Corresponding authors

E-mail address: koshinuma.terunari@aist.go.jp (T. Koshinuma), iyo-akira@aist.go.jp (A. Iyo).

## 1. Introduction

In recent years, the search for new superconductors comprising 5$d$ transition metals of Ir and Pt has been actively conducted because of the potential of distinctive superconducting states caused by the strong spin-orbit interaction (SOI) [1-11]. Verifications of unconventional superconductivity in such 5$d$ transition metal-based materials have also been vigorously pursued [12-14].

A Laves phase is a promising target for exploration of such materials, because there are many superconductors composed of 5$d$ transition metals. Among these, MgCu$_2$-type (C15) CaIr$_2$ and SrIr$_2$ Laves phases have been the subject of intensive experimental and theoretical studies [15-20]. In $A$Ir$_2$, the SOI has been reported to have a significant impact on the physical properties because of the changes in band dispersion, the complexity of the Fermi surface, and enhancement of electron–phonon coupling [15-17,20].

Figure 1 illustrates the crystal structure of the C15 Laves phase (a cubic system with a space group of $Fd$-3$m$), and Table 1 summarizes the lattice constants and superconducting transition temperatures ($T_c$) of $AM_2$ ($A$ = Ca, Sr, Ba; $M$ = Rh, Pd, Ir, Pt) reported thus far, including BaIr$_2$ synthesized in this study [21].

Superconductivity has been reported only for $AM_2$ ($M$ = Rh, Ir), and prior to our study, $AM_2$ had a higher $T_c$ with increasing atomic number of $A$ for each $M$. Gutowska $et$ $al$. performed phonon calculations on Sr$M_2$ ($M$ = Ir, Rh) and reported that the higher the mass of the constituent elements in $AM_2$, the stronger the electron-phonon coupling. In other words, if BaIr$_2$ can be synthesized, a $T_c$ higher than those of other $AM_2$ compounds can be expected. However, there were no experimental reports on BaIr$_2$, which motivated us to synthesize BaIr$_2$. The synthesis of BaIr$_2$ is also significant from the viewpoint of inorganic synthetic chemistry. $AM_2$ can be regarded as a structure in which atom $A$ is inserted in the space of a pyrochlore lattice formed by $M$ atoms. BaIr$_2$ does not crystallize under ambient pressure, possibly because of the size of Ba, which is too large to be in the tolerance range wherein the Laves phase is formed.

To crystallize BaIr$_2$ by reducing the size mismatch, herein, we utilize a high-pressure (HP) synthesis method. As a result, we succeeded in synthesizing BaIr$_2$, the C15 Laves phase, and confirmed its superconductivity. This study reports the synthesis, physical-property measurements, and electronic-structure calculation of BaIr$_2$.

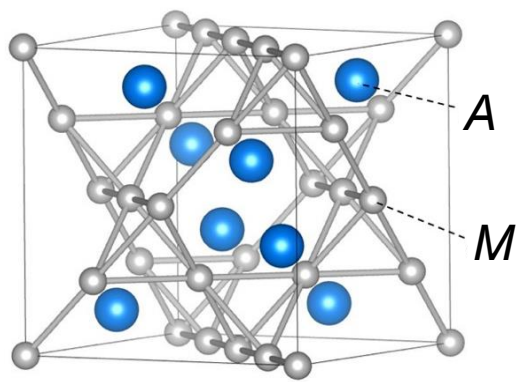

**Fig. 1** Schematic of the crystal structure of the MgCu$_2$-type Laves phase $AM_2$ (created using VESTA software [22]). Solid lines indicate the unit cell.

**Table 1**

Lattice constants and $T_c$ of $AM_2$ ($A$ = Ca, Sr, Ba; $M$ = Pd, Rh, Pt, Ir). No superconductivity has been reported in $AM_2$ ($M$ = Pd, Pt).

|    | Ca | Sr | Ba |
|----|----|----|----|
| Pd | 7.665(5) Å [23] | 7.826(5) Å [23] | 7.953(5) Å [23] |
| Pt | 7.629(5) Å [23] | 7.777(5) Å [23] | 7.920(5) Å [23] |
| Rh | 7.5326(6) Å<br>5.13 K [24] | 7.72090(5) Å<br>5.4 K [25] | 7.87163(7) Å<br>5.6 K [25] |
| Ir | 7.545 Å [23]<br>5.8 K [16] | 7.7075(2) Å<br>5.9 K [17] | 8.038(1) Å<br>2.7 K (This study) |

## 2. Materials and method

### 2.1 Sample Preparation

Polycrystalline samples were obtained via a solid-state reaction using a cubic-anvil-type HP apparatus (Oh!sawa system, CAP-035) [26]. A Ba ingot (99%, Rare Metallic Co., Ltd.) and Ir powder (99.9%, Rare Metallic Co., Ltd.) were used as starting materials. The Ba ingot was filed to form a powder. The Ba and Ir powders were weighed and mixed using a mortar in a glove box filled with $N_2$ gas. Ba in the sample decreased due to diffusion of Ba from the sample to the BN vessel during heating. To compensate for the decrease in Ba, excess Ba (5-20 mol%) was added to the starting composition. Note that metal vessels such as gold could not be used because they react with Ba. In addition, we focused on mixing the large grain-sized Ba powder (~100 μm) with the fine Ir powder (~10 μm) as uniformly as possible.

The mixed powder was pressed into a pellet and placed in a BN reaction vessel. The sample was heated for 30 min at a temperature of ~925 °C under a pressure of 3.3 GPa. BaIr$_2$ was formed at 1.9 GPa and not at 1.0 GPa, indicating that BaIr$_2$ was indeed a HP phase.

The superconductivity (transition width and volume fraction) and yield of BaIr$_2$ in the samples were strongly dependent on the excess Ba and synthesis temperature. When the synthesis temperature was low (e.g., 850 °C) or the excess Ba was too large, samples with poor superconductivity were obtained, although the yield of BaIr$_2$ in the samples was sufficiently high. It was also found that when the synthesis temperature was high (e.g., 1100 °C), samples with good superconductivity in terms of $T_c$ and transition width were obtained, but the ratio of impurity phases was relatively high. In other words, there is a trade-off between phase purity and superconductivity in polycrystalline samples, and the best balance sample was obtained at intermediate temperatures between low and high temperatures (e.g., 900 °C) with a small amount of Ba added to the synthesis. This implies that the superconductivity in BaIr$_2$ is sensitive to off-stoichiometry, as in the case of the Laves phase LuIr$_2$ [27].

The measurements conducted in this study were performed on a sample synthesized at 925 °C with 6% excess Ba. Because BaIr$_2$ reacted with $O_2$ and $CO_2$ in the air and eventually decomposed into BaCO$_3$ and Ir, the samples were treated in a glove box.

### 2.2 Physical-Property Measurements

Powder X-ray diffraction (XRD) patterns were recorded using a diffractometer (Rigaku, Ultima IV) with Cu$K\alpha$ radiation (1.5418 Å) at room temperature (~293 K). Measurements were performed using an airtight attachment that prevented the sample from being exposed to air. Rietveld fitting and simulations were performed using the RIETAN-2000 [28].

Magnetization ($M$) was measured in magnetic fields ($H$) using a magnetic-property measurement system (Quantum Design, MPMS-

XL7). Temperature (*T*), dependence of electrical resistivity (*ρ*), and specific heat (*C*) were measured using a physical-property measurement system (Quantum Design, PPMS). The composition of the samples was analyzed using an energy dispersive X-ray spectrometer (Oxford, SwiftED3000) attached to an electron microscope (Hitachi High-Technologies, TM-3000).

*2.3 Electronic-Structure Calculation*

The electronic structure of $BaIr_2$ was obtained by first-principles calculations using the full-potential linearized augmented plane wave method based on the density functional theory. The WIEN2k [29] computer program was used to conduct the entire calculation. The exchange-correlation potential was approximated using the generalized gradient approximation [30]. We used muffin-tin spheres with radii $r(Ba) = r(Ir) = 2.5$ a.u. Plane-wave cutoff $K_{max}$ was chosen such that $R_{mt}*K_{max} = 7.0$, where $R_{mt} = r(Ba) = r(Ir)$ is the radius of the smallest muffin-tin sphere. We used 1,000 and 8,000 k-points for the self-consistent field and electronic density of states (DOS) calculations. The lattice constant $a_{opt}$ was optimized to minimize the total energy and was found to be equal to 7.902 Å, which is lower than the experimental value presented later. The SOI was included in the calculations using a second variational approach.

## 3. Results and discussion

*3.1 XRD Measurement*

Figures 2(a) and 2(b) show the powder XRD pattern of the sample and the simulated diffraction pattern of the $BaIr_2$ Laves phase. They are in good agreement with each other, except for the small diffraction peaks associated with Ir and other unknown substances. Upon comparing both the XRD patterns in more detail, it was found that the relative intensity of the observed (220) peak was higher than that of the simulated (220) peak, possibly because the diffraction peak associated with the unknown substances overlaps with the (220) peak. The molar composition ratio Ba:Ir of the sample was measured to be 0.97(3):2.00(2). Based on the above experimental results, we concluded that $BaIr_2$ was obtained as the main phase in the sample.

We evaluated the lattice constant of $BaIr_2$ and the molar ratio of $BaIr_2$ and Ir in the sample using Rietveld fitting. The *a*-axis length of $BaIr_2$ was determined to be 8.038(1) Å (atomic coordinates (*x*, *y*, *z*) of Ir and Ba are (0, 0, 0) and (3/8, 3/8, 3/8), respectively), which is the longest among those for $AM_2$ listed in Table 1. The crystallization of $BaIr_2$ was possibly realized because Ba ions shrink to a greater extent than the pyrochlore lattice under HP, resulting in a tolerance range suitable for the formation of the Laves phase. The longest *a*-axis length may support the above-mentioned crystallization mechanism for $BaIr_2$. The $BaIr_2$:Ir molar ratio in the sample was estimated to be 0.81:0.19; this result was used to analyze the specific heat.

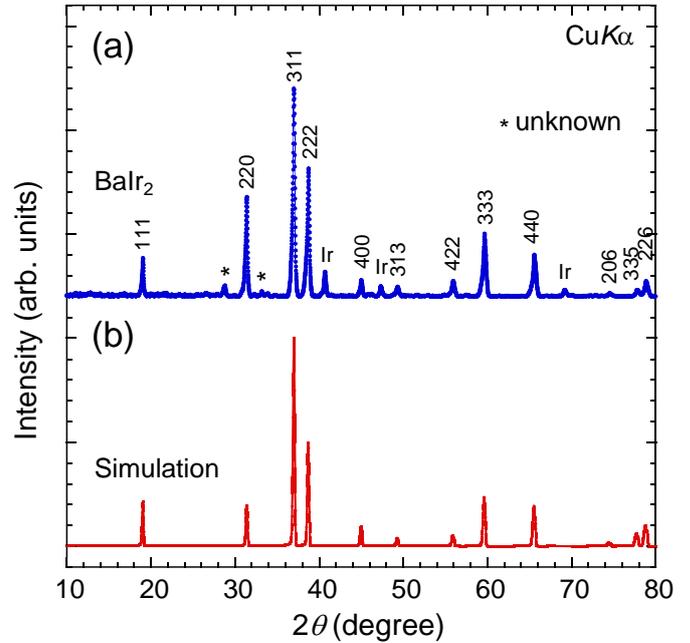

**Fig. 2**(a) Powder XRD pattern of the sample synthesized under HP. Background noise is eliminated from the data. The peaks corresponding to $BaIr_2$ are labeled with diffraction indices. (b) Simulated XRD pattern of the $BaIr_2$ assuming the $MgCu_2$-type structure.

*3.2 Superconducting Properties*

Figure 3 shows the *T* dependence of $4πM/H$ for the sample. The experimental data were demagnetization field corrected using a demagnetization factor *N* (=0.15) estimated from the shape of the measurement sample [31]. A pronounced diamagnetic transition due to the occurrence of superconductivity was observed at a $T_c$ of ~2.7 K. The zero-field-cooling (ZFC) magnetization at 2 K is 85% of the full shielding diamagnetism ($4πM/H = −1$), which was large enough as bulk superconductivity. Because Ir does not exhibit superconductivity above 2 K [32], the observed superconductivity was attributed to $BaIr_2$. The $T_c$ of $BaIr_2$ was approximately half those of $CaIr_2$ and $SrIr_2$, which is not in accordance with the theoretical prediction [15].

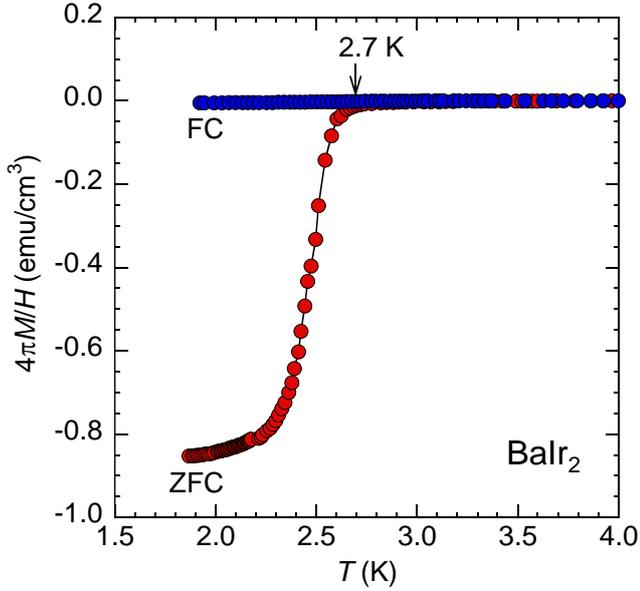

**Fig. 3** $T$ dependence of $4\pi M/H$ for the sample. Measurements were performed with ZFC and field-cooling (FC) modes under a magnetic field of 10 Oe. Demagnetization field correction was performed on $M$ considering the shape of the sample [31].

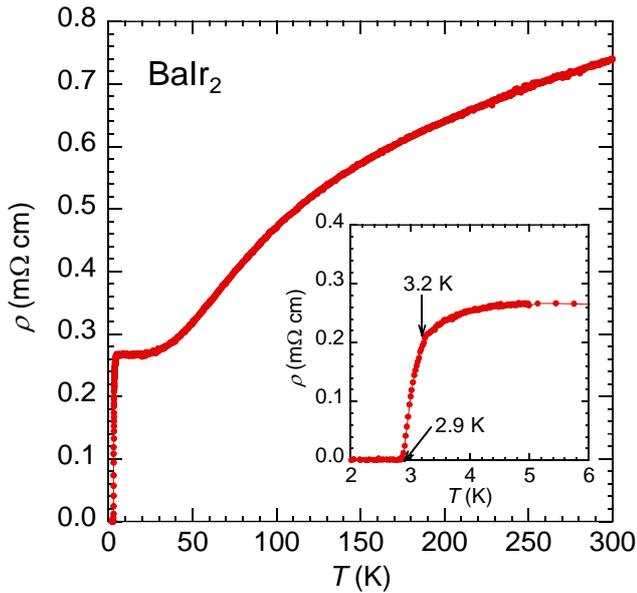

**Fig. 4** $T$ dependence of $\rho$ for the sample. The inset shows the enlargement around superconducting transition.

Figure 4 shows the $T$ dependence of $\rho$ below 300 K at $H$ = 0 Oe for the sample. The $\rho$ showed metallic $T$ dependence with a residual resistivity ratio, defined as $\rho(300\,K)/\rho(5\,K)$, of ~2.8. As shown in the inset of Fig. 4, an abrupt decrease in $\rho$ due to superconductivity was observed at ~3 K, and the resistance became zero at 2.9 K.

The $T$ dependence of $\rho$ under $H$ of up to 25 kOe for the sample is shown in Figure 5. The parallel shifts of the superconducting transition in $H$ allowed accurate evaluation of the upper critical field ($H_{c2}$).

The inset of Fig. 5 shows the $T$-dependent $H_{c2}$ defined at the midpoint of the superconducting transition curves. $H_{c2}(0)$ is obtained to be 67.7 kOe by the fitting performed with the Ginzburg–Landau (GL) theory [33], $H_{c2}(T) = H_{c2}(0)[(1-(T/T_c)^2)/(1+(T/T_c)^2)]$, where $H_{c2}(0)$ is the upper critical field at 0 K.

The $H_{c2}(0)$ of $BaIr_2$ was found to be higher than those of $CaIr_2$ and $SrIr_2$ (40–65 kOe [15-17]) despite $BaIr_2$ having a considerably lower $T_c$. Therefore, the $H_{c2}(0)$ of $BaIr_2$ exceeded the Pauli paramagnetic limit ($H_{Pauli}$) of $1.86T_c$ = 50 kOe [34,35]. The GL coherence length ($\xi_0$) was calculated to be 69.7 Å using the equation $H_{c2}(0) = \Phi_0/2\pi\xi_0^2$, where $\Phi_0$ is the magnetic flux quantum.

We estimated the orbital limit ($H_{orb}$) to be 56.2 kOe using $H_{orb}$ = $0.698T_c(|dH_{c2}/dT|_{T=T_c})$, where $T_c$ = 2.94 K and $|dH_{c2}/dT|_{T=T_c}$ = 27.4 kOe/K, based on the Werthamer-Helfand-Hohenberg (WHH) theory [36]. The Maki parameter ($\alpha_M$), which is defined as $\alpha_M = \sqrt{2}H_{orb}/H_{Pauli}$ [37], was calculated to be 1.45. This $\alpha_M$ value is larger than that expected for a typical type-II superconductor ($\alpha_M < 1$), indicating that the paramagnetic pair breaking effect is dominant in $BaIr_2$.

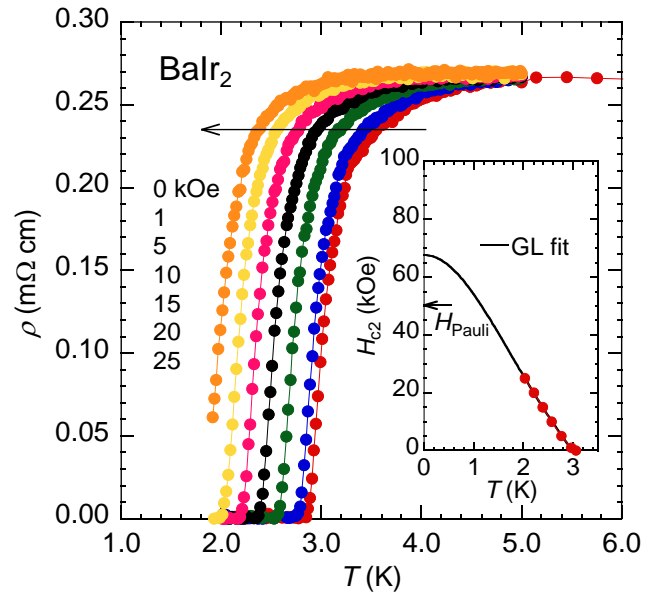

**Fig. 5** $T$ dependence of $\rho$ in $H$ of up to 25 kOe for the sample. The inset shows the $T$ dependence of $H_{c2}$ determined at the midpoint of the superconducting transition curves. The solid curve indicates the GL fitting.

Figure 6 shows the $H$ dependence of $M$ for the samples measured at several temperatures below $T_c$. At $T$ = 1.9 K, $M$ showed a broad minimum at ~260 Oe. The minimum shifted to a lower $H$ with an increase in the measured $T$.

When determining the lower critical field $H_{c1}$ based on the $M$–$H$ data, it is generally defined by the magnetic field at which the magnetization process begins to deviate from a linear behavior associated with perfect diamagnetism. However, it is probably difficult to determine the $H_{c1}$ accurately when the $T_c$ value is close to the temperature where the $M$–$H$ data was collected, as in the case of BaIr$_2$. Therefore, we defined the $H_{c1}$ by the relation $H_{c1} = H_p/(1-N)$ as reported in the literature [38]. Here, $H_p$ is referred to as the penetration field, which is defined as the intersection point between the extrapolation of perfect diamagnetism and a level of the minimum magnetization, as indicated by the dashed lines in Fig. 6. [39]. Although this method may slightly overestimate the value of $H_{c1}$, it allows for the systematic evaluation of the $H_{c1}$ at each temperature [40].

As indicated by the solid line in the inset of Fig. 6, the $H_{c1}$ thus obtained was in agreement with that reproduced by the following formula based on the GL theory: $H_{c1}(T) = H_{c1}(0)[1-(T/T_c)^2]$. We calculated $H_{c1}(0)$ to be 286 Oe. The London penetration depth $\lambda_0$ of 1520 Å was derived using the equation $H_{c1}(0) = \Phi_0/\pi\lambda_0^2$.

The GL parameter of $\kappa_{GL} = \lambda_0/\xi_0$ is determined to be 21.8, indicating that BaIr$_2$ is a type-II superconductor ($\kappa_{GL} > 1/\sqrt{2}$).

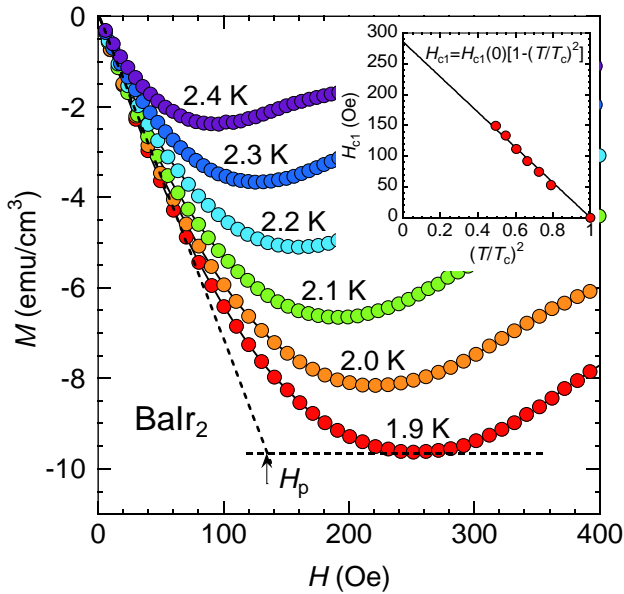

**Fig. 6** $H$ dependence of $M$ at several temperatures below $T_c$ for the sample. The inset shows $H_{c1}$ as a function of $(T/T_c)^2$. The solid line indicates a fitting based on the GL theory.

The measurement of $C$ was performed on a sample cut from the same pellet as that used for the powder XRD and magnetization measurements. The contribution of Ir in the sample to $C$ was subtracted using the reported parameters [31] according to the molar ratio in the sample estimated by Rietveld analysis [41]. The contribution of unknown substances to $C$ was not included in the analysis. However, the amount of the unknown substances in the sample was small judging from the XRD pattern; thus, the impact on the analysis results was considered to be minimal.

The $T^2$ dependence of $C/T$ for BaIr$_2$ thus obtained is shown in Figure 7. A clear jump in the specific heat indicates the bulk nature of superconductivity. The data for $T^2$ in the 7–20 K$^2$ range were fitted with $C/T = \gamma_n + \beta T^2$, where $\gamma_n$ (the Sommerfeld constant) and $\beta$ are the coefficients associated with the electron and phonon contributions to the total specific heat ($C = C_{el} + C_{ph}$), respectively. The fitting yields the values for $\gamma_n$ and $\beta$ as 12.0 mJ mol$^{-1}$ K$^{-2}$ and 1.83 mJ mol$^{-1}$ K$^{-4}$, respectively. The Debye temperature $\Theta_D$ was calculated to be 147 K using $\beta = (12/5)N\pi R\Theta_D^{-3}$, where R = 8.314 J mol$^{-1}$K$^{-2}$ and $N$ = 3 ($N$ is the number of atoms per formula unit cell). The $\Theta_D$ of BaIr$_2$ is lower than that of CaIr$_2$ and SrIr$_2$ (160–214 K [15-17]), which can be qualitatively explained as the effect of the larger atomic weight of Ba. The $T_c$ of the sample is as low as 2.7 K, and a good linear fitting is obtained at $T^2 < 20$ K$^2$, as shown in Fig 7. Therefore, the obtained values of $\gamma_n$ and $\beta$ are expected to be almost the same even if superconductivity is suppressed by applying a magnetic field.

The inset shows the $T$ dependence of $C_{el}$, which is obtained by subtracting the phonon contribution from $C$. The thermodynamically determined $T_c$ was 2.35 K. The electron–phonon coupling constant $\lambda_{e-p}$ in BaIr$_2$ was evaluated using the McMillan equation [42], $\lambda_{e-p} = (\mu^*\ln(1.45T_c/\Theta_D)-1.04)(1-0.62\mu^*)/(1.04+\ln(1.45T_c/\Theta_D))$, where $\mu^*$ is a Coulomb pseudopotential parameter. Using $T_c$ = 2.35 K, $\Theta_D$ = 147 K, and the standard $\mu^*$ value (0.13), $\lambda_{e-p}$ was calculated to be 0.63. The $\lambda_{e-p}$ of BaIr$_2$ was considerably smaller than that (0.79–1.17) of CaIr$_2$ and SrIr$_2$ [15-17]. The normalized specific heat jump ($\Delta C_{el}/\gamma_n T_c$) was estimated to be ~1.2, which is obtained from the fitting with the $\alpha$-model [43]. This value is close to that obtained through the Bardeen–Cooper–Schrieffer (BCS) model (1.43). The measured values of $\lambda_{e-p}$ and $\Delta C_{el}/\gamma_n T_c$ indicate that BaIr$_2$ is a weak-coupling superconductor. The values of $\Delta C_{el}/\gamma_n T_c$ strongly depend on $A$ (= Ca, Sr, Ba) in $A$Ir$_2$. These values were reported to be 0.89 for CaIr$_2$ and 1.71 and 2.08 for SrIr$_2$ [15-17]. The physical parameters experimentally evaluated for BaIr$_2$ in this study are listed in Table 2.

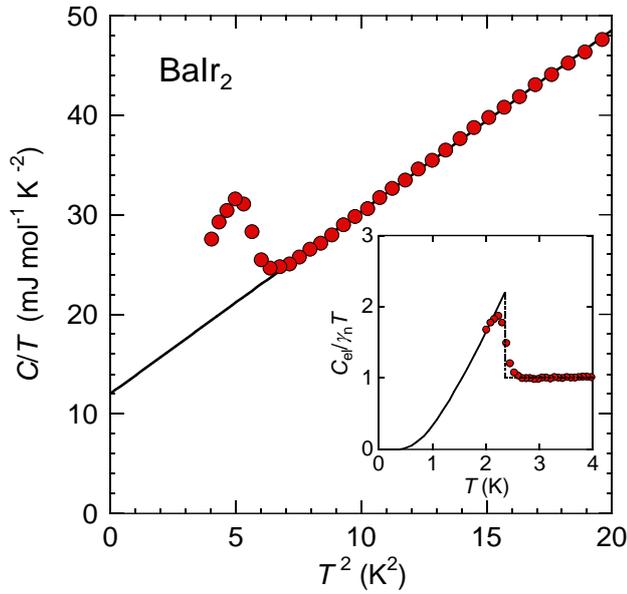

**Fig. 7** $C/T$ vs. $T^2$ for BaIr$_2$ below 5 K at the zero field. The solid line shows the fitting of data for $T^2$ = 7–20 K$^2$ with $C/T = \gamma_n + \beta T^2$ and its extrapolation to $T$ = 0 K. The inset shows the $T$ dependence of $C_{el}/\gamma_n T$. The solid curve shows the fitting with the $\alpha$-model.

**Table 2**

The physical parameters experimentally evaluated for BaIr$_2$ in this study.

| Parameters | Values |
|---|---|
| $a$ | 8.038(1) Å |
| $T_c$ | 2.7 K |
| $H_{c1}(0)$ | 286 Oe |
| $H_{c2}(0)$ | 67.7 kOe |
| $\alpha_M$ | 1.45 |
| $\lambda_0$ | 1520 Å |
| $\xi_0$ | 69.7 Å |
| $\kappa_{GL}$ | 21.8 |
| $\gamma_n$ | 12.0 mJ mol$^{-1}$ K$^{-2}$ |
| $\beta$ | 1.83 mJ mol$^{-1}$ K$^{-4}$ |
| $\Theta_D$ | 147 K |
| $\lambda_{e-p}$ | 0.63 |
| $N(0)$ | 3.12 states eV$^{-1}$ f.u.$^{-1}$ |
| $\Delta C_{el}/\gamma_n T_c$ | 1.2 |

*3.3 Electronic-Structure Calculation*

Figure 8 shows the DOS of BaIr$_2$ obtained by the first-principles calculations with the SOI. The inset shows the magnification of the total DOS near $E_F$ calculated with and without SOI. As in the cases of CaIr$_2$ [16] and SrIr$_2$ [17], the DOS at the Fermi energy ($E_F$), $N(0)$, was mostly contributed by the Ir-5$d$ states, and $E_F$ was located at a valley caused by the splitting of the DOS peak due to the strong SOI. The $N(0)^{cal}$ determined by the calculation with SOI was 3.15 states eV$^{-1}$ f.u.$^{-1}$ (f.u. is formula unit), which is in good agreement with the $N(0)$ (3.12 states eV$^{-1}$ f.u.$^{-1}$) estimated from the experimentally obtained parameters $\gamma_n$ and $\lambda_{e-p}$ using the expression $3\gamma_n = \pi^2 k_B^2 N(0)(1+\lambda_{e-p})$.

The $N(0)$ of BaIr$_2$ was comparable to that reported for CaIr$_2$ and SrIr$_2$ (3.49 and 2.74 eV$^{-1}$ f.u.$^{-1}$, respectively). Therefore, the lower $T_c$ in BaIr$_2$, approximately half of that of CaIr$_2$ and SrIr$_2$, cannot be explained by $N(0)$. The lower $T_c$ in BaIr$_2$ is possibly attributed to weaker electron–phonon interactions.

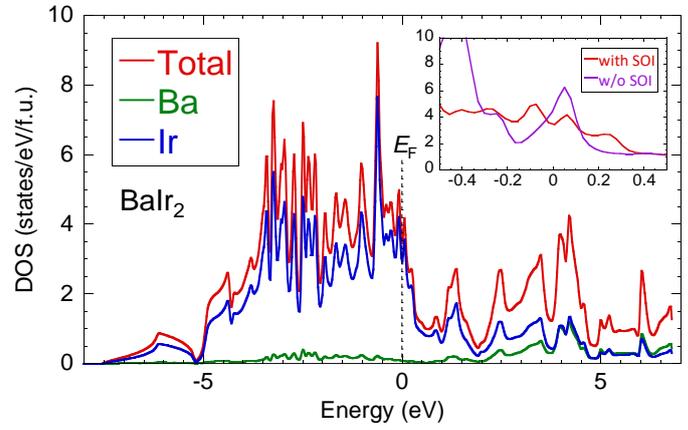

**Fig. 8** Total and partial electronic DOS for BaIr$_2$ obtained by the first-principles calculation with SOI; $E$ = 0 eV corresponds to the Fermi energy $E_F$. The inset is the magnification of the total DOS near the $E_F$ calculated with/without SOI.

**4. Conclusion**

We successfully synthesized BaIr$_2$, a novel C15 Laves phase superconductor, using the HP synthesis technique. BaIr$_2$ was found to be a type-II superconductor ($T_c$ = 2.7 K) with a large $H_{c2}(0)$ of 67.7 kOe beyond the Pauli paramagnetic limit. BaIr$_2$ was classified as a weak-coupling superconductor based on the experimentally determined $\Delta C_{el}/\gamma_n T_c$ and $\lambda_{e-p}$. The band structure calculation confirmed that the effect of the SOI on $N(0)$ in BaIr$_2$ is strong, similar to that observed in the cases of CaIr$_2$ and SrIr$_2$. In fact, the calculated $N(0)$ with the SOI was in good agreement with the experimentally estimated value. This study also revealed that the superconductivity of BaIr$_2$, such as $T_c$, $H_{c2}$, and specific heat jump, is significantly different from those of CaIr$_2$ and SrIr$_2$. Further theoretical and experimental studies are needed to elucidate the mechanism underlying the particularity of BaIr$_2$.

**ACKNOWLEDGMENTS**

This work was supported by JSPS KAKENHI (grant numbers JP19K04481, JP19K03731, and JP19H05823).

**References**

[1]. E. Bauer, G. Hilscher, H. Michor, C. Paul, E. W. Scheidt, A. Gribanov, Y Seropegin, H. Noë̈l, M. Sigrist, and P. Rogl, Phys. Rev. Lett. **92**, (2004), 027003. Heavy Fermion Superconductivity and Magnetic Order in Noncentrosymmetric CePt$_3$Si. https://doi.org/10.1103/PhysRevLett.92.027003

[2]. P. Badica, T. Kondo, K. Togano, J. Phys. Soc. Jpn. **74**, (2005), 1014-1019. Superconductivity in a New Pseudo-Binary Li$_2$B(Pd$_{1-x}$Pt$_x$)$_3$ (x = 0-1) Boride System. https://doi.org/10.1143/JPSJ.74.1014

[3]. T. Shibayama, M. Nohara, H. A. Katori, Y. Okamoto, Z. Hiroi, H. Takagi, J. Phys. Soc. Jpn. **76**, (2007), 073708. Superconductivity in Rh$_2$Ga$_9$ and Ir$_2$Ga$_9$ without Inversion Symmetry. https://doi.org/10.1143/JPSJ.76.073708

[4]. Y. Nishikubo, K. Kudo, M. Nohara, J. Phys. Soc. Jpn. **80**, (2011), 055002. Superconductivity in the Honeycomb-Lattice Pnictide SrPtAs. http://dx.doi.org/10.1143/JPSJ.80.055002

[5]. G. Eguchi, D. C. Peets, M. Kriener, Y. Maeno, E. Nishibori, Y. Kumazawa, K. Banno, S. Maki, H. Sawa, Phys. Rev. B **83**, (2011), 024512. Crystallographic and superconducting properties of the fully gapped noncentrosymmetric 5$d$-electron superconductors Ca$M$Si$_3$ ($M$ = Ir, Pt). https://doi.org/10.1103/PhysRevB.83.024512

[6]. T. Takayama, K. Kuwano, D. Hirai, Y. Katsura, A. Yamamoto, H. Takagi, Phys. Rev. Lett. **108**, (2012), 237001. Strong Coupling Superconductivity at 8.4 K in an Antiperovskite Phosphide SrPt$_3$P. https://doi.org/10.1103/PhysRevLett.108.237001

[7]. S. Pyon, K. Kudo, M. Nohara, J. Phys. Soc. Jpn. **81**, (2012), 053701. Superconductivity Induced by Bond Breaking in the Triangular Lattice of IrTe$_2$, https://doi.org/10.1143/JPSJ.81.053701

[8]. S. Pyon, K. Kudo, J. Matsumura, H. Ishii, G. Matsuo, M. Nohara, H. Hojo, K. Oka, M. Azuma, V. O. Garlea, K. Kodama, S. Shamoto, J. Phys. Soc. Jpn. **83**, (2014), 093706. Superconductivity in Noncentrosymmetric Iridium Silicide Li$_2$IrSi$_3$, http://dx.doi.org/10.7566/JPSJ.83.093706

[9]. Y. Okamoto, T. Inohara, Y. Yamakawa, A. Yamakage, K. Takenaka, J. Phys. Soc. Jpn. **85**, (2016), 013704. Superconductivity in the Hexagonal Ternary Phosphide ScIrP. https://doi.org/10.7566/JPSJ.85.013704

[10]. K. Horigane, K. Takeuchi, D. Hyakumura, R. Horie, T. Sato, T. Muranaka, K. Kawashima, H. Ishii, Y. Kubozono, S. Orimo, *New J. Phys.* **21**, (2019), 093056. Superconductivity in a new layered triangular-lattice system Li$_2$IrSi$_2$. https://doi.org/10.1088/1367-2630/ab4159

[11]. K. Kudo, H. Hiiragi, T. Honda, K. Fujimura, H. Idei, M. Nohara, J. Phys. Soc. Jpn. **89**, (2020), 013701. Superconductivity in Mg$_2$Ir$_3$Si: A Fully Ordered Laves Phase. https://doi.org/10.7566/JPSJ.89.013701

[12]. E. Bauer, and Manfred Sigrist, eds. Non-centrosymmetric superconductors: introduction and overview. Vol. 847. Springer Science & Business Media, 2012.

[13]. M. Nishiyama, Y. Inada, and Guo-qing Zheng, Phys. Rev. Lett. **98**, (2007), 047002. Spin Triplet Superconducting State due to Broken Inversion Symmetry in Li$_2$Pt$_3$B, https://doi.org/10.1103/PhysRevLett.98.047002

[14]. P. K. Biswas, H. Luetkens, T. Neupert, T. Stürzer, C. Baines, G. Pascua, A. P. Schnyder, M. H. Fischer, J. Goryo, M. R. Lees, H. Maeter, F. Brückner, H.-H. Klauss, M. Nicklas, P. J. Baker, A. D. Hillier, M. Sigrist, A. Amato, D. Johrendt, Phys. Rev B **87**, (2013), 180503(R). Evidence for superconductivity with broken time-reversal symmetry in locally noncentrosymmetric SrPtAs, https://doi.org/10.1103/PhysRevB.87.180503

[15]. S. Gutowska, K. Górnicka, P. Wójcik, T. Klimczuk, B. Wiendlocha, Phys. Rev. B **104**, (2021), 054505. Strong-coupling superconductivity of SrIr$_2$ and SrRh$_2$: Phonon engineering of metallic Ir and Rh. https://doi.org/10.1103/PhysRevB.104.054505

[16]. N. Haldolaarachchige, Q. Gibson, L. M. Schoop, H. Luo, R. J. Cava, *J. Phys.: Condens. Matter* **27**, (2015), 185701.

Characterization of the heavy metal pyrochlore lattice superconductor $CaIr_2$. https://doi.org/10.1088/0953-8984/27/18/185701

[17]. R. Horie, K. Horigane, S. Nishiyama, M. Akimitsu, K. Kobayashi, S. Onari, T. Kambe, Y. Kubozono, J. Akimitsu, *J. Phys.: Condens. Matter* **32**, (2020), 175703. Superconductivity in $5d$ transition metal Laves phase $SrIr_2$. https://doi.org/10.1088/1361-648X/ab6a2e

[18]. X. Yang, H. Li, T. He, T. Taguchi, Y. Wang, H. Goto, R. Eguchi, R. Horie, K. Horigane, K. Kobayashi, J. Akimitsu, H. Ishii, Y. Liao, H. Yamaoka, Y. Kubozono, *J. Phys.: Condens. Matter* **32**, (2020), 025704. Superconducting behavior of a new metal iridate compound, $SrIr_2$, under pressure. https://doi.org/10.1088/1361-648X/ab4605

[19]. Y. Zhang, X. M. Tao, M.Q. Tan, *Chinese Phys. B* **26** (2017), 047401. Density-functional theory study on the electronic properties of Laves phase superconductor $CaIr_2$. https://doi.org/10.1088/1674-1056/26/4/047401

[20]. H. M. Tütüncü, H. Y. Uzunok, Karaca, E. Arslan, G. P. Srivastava, *Phys. Rev. B* **96**, (2017), 134514. Effects of spin-orbit coupling on the electron-phonon superconductivity in the cubic Laves-phase compounds $CaIr_2$ and $CaRh_2$. https://doi.org/10.1103/PhysRevB.96.134514

[21]. It is mentioned in ref. [18] that the research on superconductivity of $BaIr_2$ is in progress and a paper is in preparation, but we could not find the corresponding paper.

[22]. K. Momma, F. Izumi, *J. Appl. Crystallogr.* **41**, (2008), 653-658. VESTA: a three-dimensional visualization system for electronic and structural analysis. https://doi.org/10.1107/S0021889808012016

[23]. E. A. Wood, V. B. Compton, *Acta Crystallogr.* **11**, (1958), 429-433. Laves-phase compounds of alkaline earths and noble metals. https://doi.org/10.1107/S0365110X58001134

[24]. K. Go'rnicka, R. J. Cava, T. Klimczuk, *J. Alloys Compd.* **793**, (2019), 393-399. The electronic characterization of the cubic Laves-phase superconductor $CaRh_2$. https://doi.org/10.1016/j.jallcom.2019.04.199

[25]. C. Gong, Q. Wang, S. Wang, H. Lei, *J. Phys.: Condens. Matter* **32**, (2020), 295601. Superconducting properties of $MgCu_2$-type Laves phase compounds $SrRh_2$ and $BaRh_2$. https://doi.org/10.1088/1361-648X/ab7c12

[26]. P. M. Shirage, K. Miyazawa, M. Ishikado, K. Kihou, C. H. Lee, N. Takeshita, H. Matsuhata, R. Kumai, Y. Tomioka, T. Kito, H. Eisaki, S. Shamoto, A. Iyo, *Physica C* **469**, (2009), 9. High-pressure synthesis and physical properties of new iron (nickel)-based superconductors. https://doi.org/10.1016/j.physc.2009.03.027

[27]. Y. Takano, H. Takigami, K. Ohhata, K. Sekizawa, *Solid State Commun.* **61**, (1987), 611. Superconductivity in the Lu-Ir alloy system. https://doi.org/10.1016/0038-1098(87)90371-1

[28]. F. Izumi, T. Ikeda, *Materials Science Forum* 321, (2000), 198-205. A Rietveld-Analysis Programm RIETAN-98 and its Applications to Zeolites. https://doi.org/10.4028/www.scientific.net/MSF.321-324.198

[29]. P. Blaha, K. Schwarz, G. K. H. Madsen, D. Kvasnicka, J. Luitz, WIEN2k, An Augmented Plane Wave + Local Orbitals Program for Calculating Crystal Properties (Wienna: Vienna University of Technology, 2001).

[30]. J. P. Perdew, K. Burke, M. Ernzerhof, *Phys. Rev. Lett.* **77**, (1996), 3865. Generalized Gradient Approximation Made Simple. https://doi.org/10.1103/PhysRevLett.77.3865

[31]. D. U. Gubser, R. J. Soulen, *J. Low Temp. Phys.* **13**, (1973), 211-226. Thermodynamic properties of superconducting iridium. https://doi.org/10.1007/BF00654062

[32]. R. Prozorov, V. G. Kogan, *Phys. Rev. Applied* **10**, (2018), 014030. Effective Demagnetizing Factors of Diamagnetic Samples of Various Shapes. https://doi.org/10.1103/PhysRevApplied.10.014030

[33]. J. P. Carbotte, *Rev. Mod. Phys.* **62**, (1990), 1027. Properties of boson-exchange superconductors. https://doi.org/10.1103/RevModPhys.62.1027

[34]. A. M. Clogston, *Phys. Rev. Lett.* **9**, (1962), 266. Upper Limit for the Critical Field in Hard Superconductors. https://doi.org/10.1103/PhysRevLett.9.266

[35]. B. S. Chandrasekhar, *Appl. Phys. Lett.* **1**, (1962), 7. A NOTE ON THE MAXIMUM CRITICAL FIELD OF HIGH-FIELD SUPERCONDUCTORS. https://doi.org/10.1063/1.1777362

[36]. N. R. Werthamer, E. Helfand and P. C. Hohenberg, *Phys. Rev.* **147**, (1966) 295. Temperature and Purity Dependence of the Superconducting Critical Field, $H_{c2}$. III. Electron Spin and Spin-Orbit Effects. https://doi.org/10.1103/PhysRev.147.295

[37]. K. Maki and T. Tsuneto, *Progress of Theoretical Physics* **31**, 6, (1964), 945-956. Pauli Paramagnetism and Superconducting

State. https://doi.org/10.1143/PTP.31.945

[38]. H. Kiessig, U. Essmann, W. Wiethaup, Phys. Lett. A, **71**, (1979), 467-470. Anisotropy of $H_{c1}$ and $B_0$ in superconducting niobium. https://doi.org/10.1016/0375-9601(79)90638-8

[39]. J. A. Osborn, Phys. Rev. **67**, (1945), 351. Demagnetizing Factors of the General Ellipsoid. https://doi.org/10.1103/PhysRev.67.351

[40]. Y. Tomioka, S. Ishida, M. Nakajima, T. Ito, H. Kito, A. Iyo, H. Eisaki, S. Uchida, Phys. Rev. B **79**, (2009), 132506. Three-dimensional nature of normal and superconducting states in $BaNi_2P_2$ single crystals with the $ThCr_2Si_2$-type structure. https://doi.org/10.1103/PhysRevB.79.132506

[41]. M. Fukuma, K. Kawashima, M. Maruyama, J. Akimitsu, J. Phys. Soc. Jpn. **80**, (2011), 024702. Superconductivity in $W_5SiB_2$ with the $T_2$ Phase Structure. https://doi.org/10.1143/JPSJ.80.024702

[42]. W. L. McMillan, Phys. Rev. **167**, 331 (1968). Transition Temperature of Strong-Coupled Superconductors. https://doi.org/10.1103/PhysRev.167.331

[43]. David C Johnston, *Supercond. Sci. Technol.* **26**, (2013), 115011. Elaboration of the α-model derived from the BCS theory of superconductivity. http://dx.doi.org/10.1088/0953-2048/26/11/115011